\documentclass[onecolumn,showpacs,superscriptaddress,aps,pre,twocolumn]{revtex4-2}

\usepackage{graphicx}
\usepackage{amsmath}
\usepackage{amsfonts}
\usepackage{dcolumn}
\usepackage{bm}
\usepackage{soul}
\usepackage[normalem]{ulem}
\usepackage{color}
\usepackage[caption=false]{subfig}
\usepackage[T1]{fontenc}
\usepackage{float}
\usepackage[12h=true]{scrtime}
\usepackage[dvipsnames]{xcolor}
\usepackage{cancel}
\newcommand{\B}[1]{{\bm{#1}}}

\begin{document}
\title{Analytic theory of shear localization in amorphous solids confined by Couette geometry}
%%%%%%%%%%%%%%%%%%%%%%%%%%%%%%%%%%%%%%%%%%%%%%%%%%%%%%%%%%%%%%%%%%%%%%%%%%% 
\author{Yang Fu}
\affiliation{Hangzhou International Innovation Institute, Beihang University, Hangzhou 311115, China}
\affiliation{Institute of Theoretical Physics, Chinese Academy of Sciences, Beijing 100190, China}
\author{Yuliang Jin}
\affiliation{Institute of Theoretical Physics, Chinese Academy of Sciences, Beijing 100190, China}
\affiliation{School of Physical Sciences, University of Chinese Academy of Sciences, Beijing 100049, China}
\affiliation{Center for Theoretical Interdisciplinary Sciences, Wenzhou Institute, University of Chinese Academy of Sciences, Wenzhou, Zhejiang 325001, China}
\author{Itamar Procaccia}
\email{Itamar.Procaccia@gmail.com}
\affiliation{Sino-Europe Complex Science Center, School of Mathematics, North University of China, Shanxi, Taiyuan 030051, China.}
\affiliation{Hangzhou International Innovation Institute, Beihang University, Hangzhou 311115, China}
\affiliation{Department of Chemical Physics, The Weizmann Institute of Science, Rehovot 76100, Israel}

\date{\today}

%%%%%%%%%%%%%%%%%%%%%%%%%%%%%%%%%%%%%%%%%%%%%
\begin{abstract}
``Couette geometry'' refers to two concentric rings in 2-dimensions  (or cylinders in 3-dimensions with a medium in between). 
Typically the inner and outer rings (or cylinders) rotate at different rates and the response of the medium is studied. Here we study a medium which is a two-dimensional amorphous solid, and we rotate the inner ring quasi-statically. As stress accumulates, plastic avalanches can result in shear localization, characterized by adjacent parts of the system rotating in opposite directions, 
with the maximum shear localized between them.  We derive an analytic theory that describes and explains the shear localization, providing a-priori predictions for the angle-averaged displacement field associated with the plastic drops and the shear localization.   
\end{abstract}
%%%%%%%%%%%%%%%%
%%%%%%%%%%%%%%%%
\maketitle
%%%%%%%%%%%%%%%
Shear localization and shear banding are prevalent phenomena observed in shear strained amorphous solids \cite{80Poi,03VBBB,12DHP,13DHP,13DGMPS}. In metallic glasses, for example, shear banding is a ubiquitous response to accumulated shear stress, often leading to catastrophic material failure \cite{02LSH,14MSAV}. 
Typically shear localization results from plastic events (avalanches), and thus pure elasticity theory is unable to predict and describe the phenomenon. In spite of intense research and modeling, presently there is no theory that can provide a detailed displacement field (where particles move during the stress drop including a plastic avalanche).
It is known from experimental results that shear localization can occur either near a moving boundary in a Couette geometry or in the bulk of the system \cite{00LBLG,01DTM,04FVV,04FMH}.  The aim of this paper is to provide such a theory for amorphous solids that are loaded by quasi-static shear strain. We will see that localization can occur near a boundary or in the bulk and that the theory accounts for both possibilities. 
Note that we do not study here shear banding, a phenomenon that typically occurs at yielding, separating solid and plastic-flow states, and is associated with a large-scale, system-spanning plastic event.  Shear localization occurs during smaller-scale plastic events that can occur both before and after yielding. In this study, we focus on shear localization and leave shear banding for future studies - the latter requires a theory incorporating strong nonlinear effects.

The system under study is shown in Panel (a) of Fig.~\ref{Fig1}. Full details of the model and how it is simulated are found in Appendix \ref{model}.
%%%%%%%%%%%%%%%%%%%%%%%%%%%%%%%%%%%%%%%%%%%%%%%%%%%%%% 
\begin{figure}[h!]
    \includegraphics[width=0.6\linewidth]{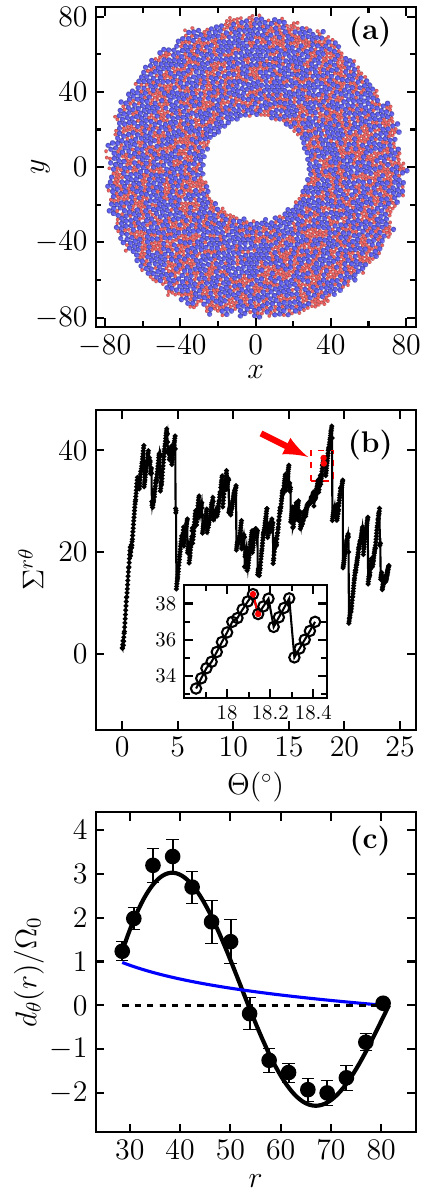}
\caption{Panel (a): The simulated system, here the area fraction is $\phi=0.91$, see text for details. Panel (b): typical shear stress vs. accumulated rotation angle $\Theta$. The stress drop responsible for Panel (c) is indicated by an arrow. 
Panel (c): typical profile of the angle-averaged angular component of the displacement field $d_\theta(r)$, normalized by inner rotation $\Omega_0$, as a function of $r$, showing the reversal of particle displacement from anti-clockwise to clockwise. This leads to the stress localization shown in Fig.~\ref{loc}.
The blue curve represents the tangential displacement of the elastic solution shown in Eq.~(\ref{d2}). For additional comparisons between theory and simulations see Fig.~\ref{average} and Fig.~\ref{loc4} in Appendix D.}
\label{Fig1}
\end{figure}
%%%%%%%%%%%%%%%%%%%%%%%%%%%%%%%%%%%%%%%%%%%%%
In short, the material is made of assemblies of $N=3534$ bidisperse 50:50 repulsive disks of radii 1 and 1.4 respectively. Below, the smaller radius is used as the unit of length. 
The material is bounded between two concentric rings of radii $R_{\rm in} =28$ and $R_{\rm out}=80.8$. The particles adjacent to the outer boundary are glued to the boundary, dictating a boundary condition $\B{d}(R_{\rm out})=\B{0}$ where $\B{d}$ is the displacement field. 
An initial random configuration of a chosen area fraction $\phi$ is prepared, and subsequently brought to mechanical equilibrium by molecular dynamics with appropriate damping.  For the present study we do not attempt to anneal further \cite{15GKPP,20SOB}, we are interested in ductile response \cite{15BPDB}. The material is then loaded by a quasi-static protocol in which the inner ring is rotated anti-clockwise in steps of $\delta \theta =0.024^{\circ}$; after each step the system is equilibrated by damped molecular dynamics.  This imposes the inner boundary condition $d_\theta(R_{\rm in})=\Omega_0=R_{\rm in} \delta \theta \pi/180\approx 0.012$ and $d_r(R_{\rm in})=0$ for the tangential and radial components of the displacement field respectively. 

Initially the accumulated shear stress $\Sigma^{r\theta}$ is increasing via mostly purely elastic steps. At higher values of stress, the response to a small additional rotation step can be a sharp stress drop due to plastic events. 
We are interested in the displacement field that is associated with such stress drops,
\begin{equation}\label{L3}
	\B {d} = d_{r}(r,\theta)\hat{r} + d_{\theta}(r,\theta)\hat{\theta}.
\end{equation}
%%%%%%%%%%%%%%%%
Since the displacement field is different in every realization, we will focus on an analytic theory for the angle-averaged tangential and radial component, 
\begin{equation}	
	d_{\theta}(r)\equiv \frac{1} {2\pi}\! \int_0^{2\pi}d_{\theta}(r,\theta) d\theta, \ \quad 	d_{r}(r)\equiv  \frac{1} {2\pi} \!\!\int_0^{2\pi}d_{r}(r,\theta) d\theta\ . 
	\end{equation} 
We learn that this angle averaged $r$-dependent displacement fields are predictable by analytic theory in good agreement with numerical simulations. A typical plot of stress vs. accumulated angle is shown in Fig.~\ref{Fig1} Panel (b). As explained, this plot is characteristic of ductile materials. The profile of $d_{\theta}(r)$ which is associated with the stress drop that is marked in the figure of Panel (b) is shown in Panel (c). We note that because of the double zero boundary condition on the radial component of the displacement field, this component is negligible throughout this discussion.
The data in the Panel (c) of Fig.~\ref{Fig1} indicates that part of the system is displaced clockwise and a part anti-clockwise. This is in clear variance with the prediction of classical elasticity theory \cite{Landau} which is shown as the blue curve in panel (c). 

If elasticity theory were appropriate for describing our simulations, the classical equation for the displacement field would have read
\begin{equation}
	\mu  \Delta \B{d} + \left(\lambda + \mu \right) \nabla \left(\nabla\cdot \B{d}\right) = 0 \ , \quad\text{purely elastic.}
	\label{d1}
\end{equation}
Here $\lambda$ and $\mu$ are the classical Lam\'e coefficients. Solving Eq.~(\ref{d1}) for our system with the appropriate boundary conditions, we would expect a tangential displacement field that reads
\begin{equation}
	d_{\theta}(r)= \Omega_0 \frac{R_{\rm in}(R_{\rm out}^2-r^2 )}{r(R_{\rm out}^2-R_{\rm in}^2)}.
	\label{d2}
\end{equation} 
This exact elastic solution is shown as the blue line in Fig.~\ref{Fig1} Panel (c).
Evidently, classical elasticity theory fails miserably in reproducing, let alone predicting, the actual displacement field. 
Below we present a theory that results in the black curve in Panel (c). We will show that the theory can predict the actual displacement fields that are found in the present simulation. We use this function (which is fully derived below) to compute the shear strain as $u_{r\theta}\equiv (\partial_\theta d_r+\partial_r d_\theta)/2$. Realizing that the first term is negligible, we obtain the shear strain that is plotted in Fig.~\ref{loc}. Other examples of similar results in different realizations are shown in Appendix D. The realized displacement field and the shear localization that is seen as the minimum in this curve are the phenomena that we need to explain.
%%%%%%%%%%%%%%%%%%%%%%%%%%%%%%%%%%%%%%%%%%%%%%%%%%%%%% 
\begin{figure}[h!]
    \includegraphics[width=0.9\linewidth]{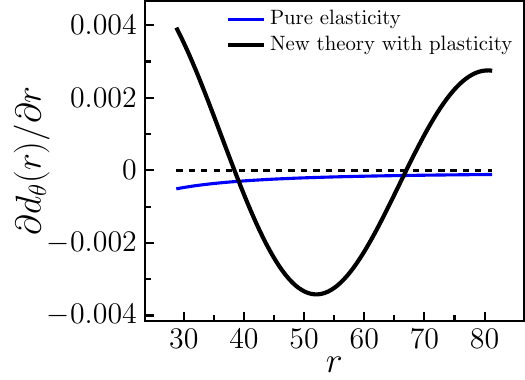}
	\caption{The shear strain as a function of $r$, exhibiting the shear localization at $r\approx 108$ where the displacement field in Panel (c) of Fig.~\ref{Fig1} changes sign. The blue curve represents the shear strain of the elastic solution. }
	\label{loc}
	\end{figure}
%%%%%%%%%%%%%%%%%%%%%%%%%%%%%%%%%%%%%%%%%%%%%%%%%

In a recent series of publications, being theoretical \cite{21LMMPRS,23CMP,24KP,24JPS},  experimental \cite{22MMPRSZ,25CSWDM} and simulational \cite{22BMP,22KMPS,23MMPR}, it was shown already that to explain generic responses of amorphous solids to external loads one needs to take into account the plastic events that appear as topological quadrupoles and dipoles in the displacement field. 
The present problem requires an extension, since we explore here not a single step of load, but a quasi-static protocol that creates a substantial background stress $\Sigma^{r\theta}$ which was not taken into account previously. 
In order to do it, here we will use the fact that in quasi-static protocols, prior to the studied stress drop the system underwent energy minimization, reached a mechanical equilibrium state with $\partial_\alpha \Sigma^{\alpha\beta}=0$. Since the next step of strain $\delta \theta =0.024^{\circ}$ is so small, we can assume that the resulting displacement field which is associated with the stress drop is sufficiently small to justify the linear approximations $u_{\alpha\beta} =(\partial_\alpha d_\beta+\partial_\beta d_\alpha)/2$ for the step of strain and $\sigma^{\alpha\beta} =A^{\alpha\beta\gamma\delta} u_{\gamma\delta}$ for the associated stress, where $A^{\alpha\beta\gamma\delta}$ is the elastic tensor that is relevant at the present state of the system. 
Note that here $\Sigma^{\alpha \beta}$ is the accumulated background stress, and $\sigma^{\alpha \beta}$ is the small change of stress during the plastic event.

As in previous studies, it is advisable to consider first the influence of quadrupoles that are formed due to plastic events, and in a second step the dipolar contributions that are formed due to gradients in the quadrupolar fields. We therefore write the Lagrangian including the costs of quadrupoles and quadrupole-quadrupole interactions in the form
\begin{eqnarray}
&&	U =\frac{1}{2} \int d^2x ~(\Sigma^{\alpha\beta}u_{\alpha\beta} + \sigma^{\alpha\beta}u_{\alpha\beta})\nonumber\\ &&+  \frac{1}{2} \int d^2x ~\Lambda_{\alpha\beta\gamma\delta}Q^{\alpha\beta}Q^{\gamma\delta}+ \int d^2x ~\Gamma_{\gamma\delta}^{\alpha\beta}u_{\alpha\beta}~Q^{\gamma\delta},
\end{eqnarray}
where $\Lambda_{\alpha\beta\gamma\delta}$ and $\Gamma_{\gamma\delta}^{\alpha\beta}$ are  coupling tensors that were introduced before \cite{21LMMPRS,23CMP,24KP,24JPS}.
Next we minimize $U$ with respect to the quadrupolar field $\bf Q$ and displacement field $\B{d}$ which are the fundamental fields in our problem.
In Appendix \ref{derive}, we show that the variation with respect to $\bf Q$ provides the relation of $\bf  Q$ to the strain field, 
\begin{align}\label{LQ8}
	Q^{\alpha\beta} = -\tilde{\Lambda}^{\alpha\beta\gamma\delta}~u_{\gamma\delta}, 
\end{align}
%%%%%%%%%%%%%%%%%
where $\tilde{\Lambda}^{\alpha\beta\gamma\delta} = \Lambda^{\alpha\beta\mu\nu}\Gamma_{\mu\nu}^{\gamma\delta}$, and $\Lambda^{\alpha\beta\gamma\delta}$ is the inverse of $\Lambda_{\alpha\beta\gamma\delta}$. 
The variation with respect to $\B{d}$ reads (see Appendix \ref{derive} for details)
%%%%%%%%%%
\begin{align}\label{LQ9}
	\delta_{d}U =  \int d^2x ~ (\Sigma^{\alpha\beta}+ \tilde{\sigma}^{\alpha\beta})~\delta u_{\alpha\beta}, \quad \tilde{\sigma}^{\alpha\beta} = \sigma^{\alpha\beta}  + \Gamma_{\gamma\delta}^{\alpha\beta} Q^{\gamma\delta}. 
\end{align}
%%%%%%%%%%
Writing now $\delta u_{\alpha\beta} = \frac{1}{2}(\delta d_{\alpha,\beta} + \delta d_{\beta,\alpha}  )$ and integrating by parts, and using  $\nabla \cdot\B \Sigma$=0, we find again the equation 
\begin{align}\label{LE0}
\frac{\partial \tilde\sigma_{\alpha\beta}}{\partial x_\alpha}= 0.
\end{align}
All that happened due to the quadrupoles alone is a renormalization of the elastic tensor. The explicit normalization of the elastic tensor is provided in Appendix \ref{derive}. 
%%%%%%%%%%%%%%%%%%%%%%%%%%%%%%%%%%%%%%%%%%%%%%%%%%%%%% 
\begin{figure}
	\includegraphics[width=0.8\linewidth, height=0.6\linewidth]{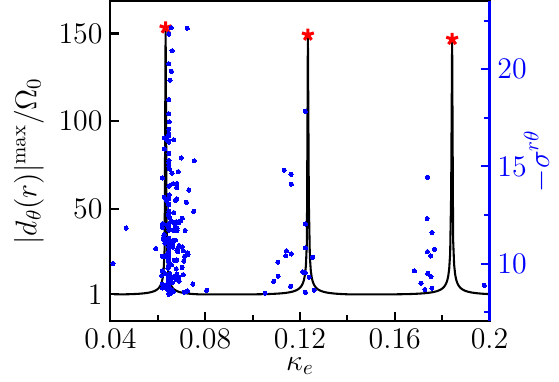}
	%\vskip 0.5 cm
	%\includegraphics[width= 7cm]{Localization.Fig3.png}
	\caption{
	Plot of the maximal value of $|d_\theta(r)|$ normalized by inner rotation $\Omega_0$ as a function of the screening parameter $\kappa_e$. The peaks represent the values of $\kappa_e$ where the denominator of Eq.~(\ref{solbess}) goes to zero, and the response of the system is maximal. The blue dots represent best fits of $\kappa_e$ to actual measured displacement fields in our simulations. The preference of the system to select responses characterized by discrete values of $\kappa_e$ is obvious.
		The height of every blue dot represents the stress drop $-\sigma^{r\theta}$ that is defined as the difference in total stress between the mechanical equilibrium configurations before and after a small rotation at the inner boundary which resulted in the fitted displacement field.}
	\label{sings}
\end{figure}
%%%%%%%%%%%%%%%%%%%%%%%%%%%%%%%%%%%%%

Once we have reached the conclusion that even in the presence of a large background stress $\bf \Sigma$ the quadrupoles only renormalize the elastic tensor, the addition of dipoles due to the gradients of the quadrupolar field $P^\alpha \equiv \partial_\beta Q^{\alpha\beta}$ follows verbatim the derivation provided in, say, Ref.~\cite{23CMP}, and see Appendix B for details. The final equation that the displacement field should satisfy reads \cite{24FHKKP}
\begin{equation}\label{L2}
	\Delta \B{d} + (1+ \tilde\lambda)\B \nabla (\B \nabla \cdot \B{d}) +\B \Gamma \B{d} =\B{0},
\end{equation}
where
\begin{equation}
	\B \Gamma=
	\begin{bmatrix}
		\kappa_{e}^{2} & \mp\kappa_o^{2}  \\
		\pm\kappa_o^{2} & \kappa_{e}^{2} 
	\end{bmatrix} \ .
	\label{odd}
\end{equation}
The $\mp$ sign of $\kappa_0^2$ means that it can be negative or positive, with an anti-symmetric counterpart $\pm \kappa_0^2$.
This equation is analytically soluble. Taking into account the periodic boundary conditions on the angle $\theta$ we can now seek a solution of these equations in the form of Fourier series for $d_{r}(r,\theta)$ and $d_{\theta}(r,\theta)$ respectively:
\begin{align}\label{L6}
	&d_{r}(r,\theta) = d_r(r) + \sum_{n=1}^{\infty}\left[  a_{n}(r)\cos(n\theta) + c_{n}(r)\sin(n\theta)\right] , \nonumber \\
	&d_{\theta}(r,\theta)  = d_\theta(r) + \sum_{n=1}^{\infty} \left[ g_{n}(r)\cos(n\theta) +e_{n}(r)\sin(n\theta)\right] .
\end{align}
%%%%%%%%%%%%
In Appendix C it is shown that the quantities of interest, i.e., $d_r(r)$ and $d_\theta(r)$ are obtained by solving the coupled equations (with a prime denoting derivative with respect to $r$):
\begin{align}\label{A00}
	[r^2d_{r}^{\prime\prime} + rd_{r}^{\prime}-d_{r} ]  +  \frac{\kappa_{e}^{2}r^2}{(\tilde\lambda +2)}  d_{r} -\frac{\kappa_o^{2}r^2}{(\tilde\lambda +2)}  d_{\theta}   & =0 ,  \nonumber \\ 
	%%%%%%%%
	[r^2d_{\theta}^{\prime\prime}  + rd_{\theta}^{\prime} -d_{\theta} ]  + \kappa_{e}^{2} r^2 d_{\theta}  +   \kappa_o^{2} r^{2} d_{r}    &=0.
\end{align}
In our case we need to satisfy the boundary conditions 
%%%%%%%%%%%%%%
\begin{align}\label{A29}
	d_{r}(r)|_{r=R_{\rm in}} = 0,  ~~~~~~~~~~~~ d_{r}(r)|_{r=R_{\rm out}} = 0, \nonumber \\
	d_{\theta}(r)|_{r=R_{\rm in}} = \Omega_{0},  ~~~~~~~~~~~~ d_{\theta}(r)|_{r=R_{\rm out}} = 0.
\end{align}
%%%%%%%%%%%%%
With these boundary conditions we find that the solution for the displacement field is, to a very good approximation, 
\begin{equation}
	\begin{aligned}
		d_{r} &= 0, 
		\\
		d_{\theta} &=\Omega_{0}\frac{J_{1}(\kappa_{e} R_{\rm out})  Y_{1}(\kappa_{e} r) - Y_{1}(\kappa_{e} R_{\rm out})  J_{1}(\kappa_{e} r)}{Y_{1}(\kappa_{e} R_{\rm in}) J_{1}(\kappa_{e} R_{\rm out}) - Y_{1}(\kappa_{e} R_{\rm out}) J_{1}(\kappa_{e} R_{\rm in})}
		%\equiv BX_{1}+(1-B)X_{2} \ ,
		\label{solbess}
	\end{aligned}
\end{equation}
where $J_1$ and $Y_1$ are the circular Bessel functions of the first
and second kind, respectively. In other words, we get away with one screening parameter, $\kappa_e$, and can safely neglect $\kappa_o$, that in the present simulation is sufficiently close to zero. The black line in Fig.~1c is the analytic form shown in Eq.~(\ref{solbess}) with $\kappa_e=0.113$ and $\kappa_o=0$. 

The great advantage of the simplified analytic form Eq.~(\ref{solbess}) is that it provides us a convenient a-priori predictability of the selected scaling exponent $\kappa_e$ in the present problem. It was shown before \cite{24JPS,25PS} that as the denominator in Eq.~(\ref{solbess}) goes to zero for discrete values of $\kappa_e$, the system's plastic response greatly prefers these values of screening parameter.The reader should note that the values of the selected $\kappa_e$ depend on the geometrical parameters $R_{\rm in}$ and $R_{\rm out}$ only, and not on the state of the system. In the present geometry more than one value is allowed, and which of those values will be realized is determined by subtle differences in the configurations. We found that out of the 200 studied drops 163 events the fitted values of $\kappa_e$ cluster around the lowest value of the predicted $\kappa_e$, and 20 events are consistent with the second value. As a rule of thumb we can state that large stress drops resulted in stress localization near the inner boundary, and smaller ones in the bulk.  A plot of the maximal value of $|d_\theta(r)|$ as a function of $\kappa_e$ is shown in Fig.~\ref{sings}. The blue dots represent best fits of $\kappa_e$ to actual measured displacement fields in the top 200 largest stress drops in 30 independent samples in our simulations.  
In all these examples the selected values of the screening exponents fall in the vicinity of those corresponding to a singularity in Eq.~(\ref{solbess}).

\begin{figure}
	\includegraphics[width=1\linewidth]{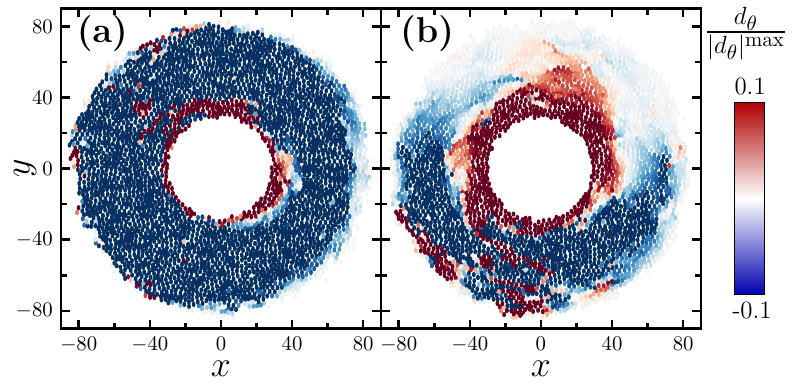}
 	\caption{Color maps of the displacement field. Red indicates anti-clockwise and blue clockwise displacement.  
    Panel (a): Typical map of displacement field when the screening exponent $\kappa_e$ is in the vicinity of $\kappa_e =0.067$. In this case the shear localization occurs near the inner boundary.
    Panel (b):  Typical map of displacement field when the screening exponent $\kappa_e$ is in the vicinity of $\kappa_e =0.123$. Here the shear localization occurs in the bulk.}
    	\label{maps}
 \end{figure}

The discrete values of $\kappa_e$ determine the analytic shape of the angle-averaged displacement field $d_\theta(r)$, and where the shear localization is located. When the selected value is in the vicinity of $\kappa_e =0.123$, the shear localization is around the middle of our Couette cell, whereas when $\kappa_e\approx 0.067$, the shear localization is adjacent to the inner ring. The difference is visually apparent in the maps of the displacement field that are shown in Fig.~\ref{maps}. One should mention that both possibilities were observed both in our simulations and in laboratory experiments \cite{00LBLG,01DTM,04FVV,04FMH}. Screening parameters in the vicinity of the third peak, at $\kappa_e\approx 0.183$ result in an angle averaged function $d_\theta(r)$ that has two maxima and one minimum, but is only rarely observed. 
 %%%%%%%%%%%%%%%%%%%%%%%%%%%%%%%%%%%%%%
%%%%%%%%%%%%%%%%%%%%%%%%%%%%%%%%%%%%%%%%%%%%%%%%%%%%%%%

\begin{figure}
	\includegraphics[width=0.7\linewidth]{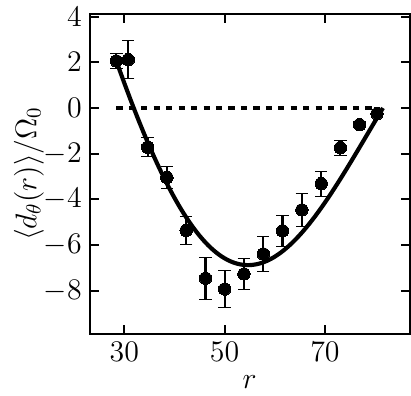}
	\caption{ Ensemble average over 10 different realizations whose displacement field is like those shown for angle averaged single realizations in panels (a) and (b)  Fig.~\ref{loc4}.} 
\label{average}
\end{figure}

 Finally, readers may ask why we present comparisons between angle averaged displacement field in theory and simulations and not ensemble averages. The first reason is that in the theory we derived an equation for angle averaged displacement field, and this is the one that we solve analytically. It is thus natural to compare it to the same quantity in simulations. The second reason is that once we angle average, additional ensemble averaging does not change much. To make this point clear we present in Fig.~\ref{average} an average over 10 different realizations of the angle averaged displacement field that are associated with simulations whose screening parameter is
in the vicinity of $\kappa_e =0.067$. Clearly, comparing to the angle averaged results of single realizations as shown in panels (a) and (b) of Fig.~\ref{loc4} in the appendix, we do not learn anything new. 
%%%%%%%%%%%%%%%%%%%%%%%%%%%%%%%%%%%%%%%%%%%

{\bf Conclusion:} 
The points to stress as a summary are the following: 
(i) We show that the angle averaged displacement field that results from the stress drop in a sheared ductile amorphous solid can be computed analytically from a theory that takes into account the topological charges that decorate the displacement field when plastic events are important. 
(ii) The theory,  which employs Lagrangians that are expanded up to quadratic order (i.e., providing linear theory for the displacement field), is applicable to quasi-static shear since each step is small and the background accumulated stress is divergence-less. It will not be appropriate when a large shear is imposed in one step. (iii) In principle one can find shear localization close to the rotating disk or in the bulk. In our quasi-static simulations the largest stress drops tended to localize near the boundary rather than in the bulk. 
(iv) In brittle materials, we expect to observe a sharp shear band associated with the yielding of the material, and this is an instability that requires a nonlinear extension of the present theory. We propose that adding quartic dipole term ($P^\alpha P^\beta P^\gamma P^\delta$) to the present Lagrangian, will be sufficient to predict and explain the creation of shear bands.
Needless to say, this extension goes beyond the scope of this paper, and will be reported in a later publication. 

{\bf Acknowledgments: } We thank Jin Shang for interesting discussions. Y.F. acknowledges funding from the China Postdoctoral Science Foundation (Grant No. 2025M784261), and the National Natural Science Foundation of China (Grant Nos. 12547175 and 22193032). IP acknowledges that this work was carried
out in part at the Sino-Europe Complex Science Center, North University of China, Taiyuan, Shanxi 030051,
China. Thanks to Profs. Guiquan Sun, Xiaofeng Luo,
for helpful discussions. Y.J. acknowledges funding from  Wenzhou Institute (Grant WIUCASICTP2022).

\appendix

\section{Model and Simulation Method}
\label{model}
\subsection{Model}
The two-dimensional model employed in our numerical simulations is based on our previous work \cite{25FYDP}, which now consists of equimolar frictionless bidisperse disks with radii $R_{1}=1$ and $R_{2}=1.4$ (the length unit is the radii of small disks).
The normal force ${\bm{F}}_{ij}$ between particles $i$ and $j$ is,
\begin{align}
	{\bm{F}}_{ij}^{(n)} = k_n \Delta_{ij}^{(n)} \hat{{\bm{n}}}_{ij} - \gamma_{n} {\bm{v}}_{ij}^{(n)},
\end{align}
where
${\bm{r}}_{ij}$ denotes the vector between the center of particle $i$ and the center of contacting particle $j$, and
$\hat{{\bm{n}}}_{ij} = {\bm{r}}_{ij}/r_{ij}$ is the unit vector of the inter-particle distance, and ${\bm{v}}_{ij}^{(n)}$ is the normal component of the relative velocity.
The overlap between two contacting particles is defined as $\Delta_{ij}^{(n)}=R_{i}+R_{j}-\left| \bm{r}_{i}-\bm{r}_{j}\right|$.
The spring coefficient is $k_n=k_n^{'} \sqrt{\Delta_{ij}^{(n)} R_{ij}}$ for the Hertzian interaction, with $k_n^{'} = 2\times 10^{5}$ and 
$R_{ij}^{-1}=R_{i}^{-1}+R_{j}^{-1}$. 
$k_n^{'}$, combined with the unit of length, determines the unit of force.
The damping coefficient is $\gamma_n = 500$.
The mass of all particles is set to be one.

\subsection{Simulation methods}
The initially random configuration is generated under periodic boundary conditions at a fixed packing fraction $\phi=0.91$. The system is then rapidly quenched to reach mechanical equilibrium. 
After this step, simulations are performed in an annular area with two boundary conditions: particles within $r<R_{\rm{in}}=28$ are chosen to quasi-statically rotate serving as the inner boundary, and particles at $r \geq R_{\rm out}=80.8$ are fixed as the outer boundary.

The quasi-static rotation simulation is carried out as follows. For each rotation step, we first choose the inner boundary particles, then rotate them anti-clockwisely in a step of $\delta \theta=0.024^{\circ}$, and then minimize the energy of $N=3534$ particles within the two concentric rings.
The choice of the inner boundary particles is fixed during the damped molecular dynamics simulations. 
The minimization procedure is terminated when the net force per particle $F_{\rm net} \leq 10^{-7}$. 
The above protocols are implemented using the LAMMPS package \cite{THOMPSON2022108171}.
The presented data are run over 30 independent samples.
Rattlers (particles with fewer than $d+1$ contacts) are removed from the mechanical equilibrium configurations before and after the rotation of inner boundary.

The stress for the every contact in the configuration is calculated based on 
\begin{align}
	{\sigma}_{ij}^{\alpha\beta} = {r_{ij,\alpha}} \otimes {F_{ij,\beta}},
\end{align}
where $\alpha=x$ or $y$ in the Cartesian coordinate, and $\otimes$ signifies the vector outer product.
The shear stress $\sigma^{r\theta}_{ij}$ for every contact in the polar coordinate is then calculated as 
\begin{align}
	\sigma^{r\theta}_{ij}=(\sigma^{yy}_{ij} -\sigma^{xx}_{ij}) \sin(2\theta)/2 + \sigma^{xy}_{ij} \cos(2\theta)
	,
\end{align}
where $\theta$ is the polar angle of particle $i$.
For the whole annulus, the particle-averaged shear stress is defined as
\begin{align}
	\sigma^{r\theta}=\sum_{i\neq j} \sigma^{r\theta}_{ij}/S,
\end{align}
where the annular area $S=\pi (R_{\rm{out}}^2-R_{\rm{in}}^2)$.

\section{Derivation of the Equation for the Displacement Field after a Stress Drop in Quasi-static Straining}
\label{derive}

\subsection{Equilibrium Equation in Classical Elasticity Theory}

The equilibrium equation for the stress is derived by minimizing the Lagrangian with respect to the virtual displacement, and reads
%%%%%%%%%%%%
\begin{align}\label{LE1}
	(\sigma^{\alpha\beta})_{,\alpha}=\frac{\partial \sigma^{\alpha\beta}}{\partial x_{\alpha}} = 0.
\end{align}
%%%%%%%%%%%
%%%%%%%%%
For an isotropic and homogeneous medium, the relation between stress and strain for the solid is %%%%%%%%%%%%%%
\begin{align}\label{LE2}
	\sigma^{\alpha\beta} = A^{\alpha\beta\gamma\delta}u_{\gamma\delta}=\lambda u_{\gamma\gamma} \delta_{\alpha\beta} + 2\mu u_{\alpha\beta},
\end{align}
%%%%%%%%%%%%%%
where $\lambda$ and $\mu$ are the Lame's constants.

With the small deformation, a linear relationship will exist between strain tensor $u_{\alpha\beta}$ and the gradients of the displacement field $d_{\alpha,\beta}$ expressed as
%%%%%%%%%
\begin{align}\label{LE3}
	u_{\alpha\beta} = \frac{1}{2} ( d_{\alpha,\beta} + d_{\beta,\alpha}  ).  
\end{align}
%%%%%%%%

Putting Eqns.(\ref{LE1}-\ref{LE3}) together, we obtain the stress equilibrium equation in terms of displacement field, i.e., the Navier-Lame equation, as
%%%%%%%%%
\begin{align}\label{LE4}
	(\lambda+\mu) u_{j,ji} + \mu u_{i,jj} =0.
	%%%%%%%%%%%
\end{align}
In terms of operators, Eqn.(\ref{LE4}) can be also written as 
\begin{equation}
	\begin{aligned}
		\mu \Delta {\bm d} + (\lambda+\mu) \nabla (\nabla \cdot {\bm d}) =0.
		\label{LE5}
	\end{aligned}
\end{equation}
which is equivalent to Eqn.(\ref{LE1}).

\subsection{Introducing Quadrupoles after Strain Drop}

In the quasi-static loading procedure, the stress is increasing on average, and when we consider a stress drop that might result in shear localization, there is a background stress, which is a state function. To write our Lagrangian, we note that strain is not a state function, and we can only focus on the strain induced by the displacement field in the last, $n$'th step.

We will denote the background stress at the ($n-1$)th step as $\bf{\Sigma}$ and write
\begin{align}
	{\bf{\Sigma}} = \sum \limits_{i=1}^{n-1} {\bm{\sigma}}_{i}.
\end{align}

Since we equilibrate after every step, $\nabla \cdot {\bf{\Sigma}}={\bm{0}}$.
In addition, the elastic tensor had been normalized in previous steps, and we employ the current tensor that is relevant for the $n$th step.

Now we write the energy functional incorporating quadrupoles (without dipoles) because quadrupoles microscopically serve as the elementary event of plasticity \cite{12DHP} and are modeled as the basic field here. The total energy of the system can be written as
\begin{align}\label{LQ1}
	U = U_{\text{el}} + U_{\text{QQ}} + U_{\text{Q-el}}. 
\end{align}
Here, $U_{\rm el}$ is the elastic energy at the $n$th step of the applied strain.
\begin{align}\label{LQ2}
	U_{\text{el}} &= \frac{1}{2} \int d^2x ~(\Sigma^{\alpha\beta}u_{\alpha\beta} + A^{\alpha\beta\gamma\delta}u_{\alpha\beta}u_{\gamma\delta})\nonumber \\
	&= \frac{1}{2} \int d^2x (\Sigma^{\alpha\beta}u_{\alpha\beta} + \sigma^{\alpha\beta}u_{\alpha\beta}),
\end{align}
%%%%%
%%%%%
\begin{align}\label{LQ3}
	U_{\text{QQ}} = \frac{1}{2} \int d^2x ~\Lambda_{\alpha\beta\gamma\delta}Q^{\alpha\beta}Q^{\gamma\delta},
\end{align}
%%%%%
%%%%%
%%%%%
\begin{align}\label{LQ4}
	U_{\text{Q-el}} =  \int d^2x ~\Gamma_{\gamma\delta}^{\alpha\beta}u_{\alpha\beta}~Q^{\gamma\delta}.
\end{align}
%%%%%
Thus $U$ is
%%%%%%%%%%
%%%%%
\begin{widetext}
\begin{align}\label{LQ5}
	U = \frac{1}{2} \int d^2x ~(\Sigma^{\alpha\beta}u_{\alpha\beta} + \sigma^{\alpha\beta}u_{\alpha\beta}) + \frac{1}{2} \int d^2x ~\Lambda_{\alpha\beta\gamma\delta}Q^{\alpha\beta}Q^{\gamma\delta} + \int d^2x ~\Gamma_{\gamma\delta}^{\alpha\beta}u_{\alpha\beta}~Q^{\gamma\delta}.
\end{align}
%%%%%
%%%%%%%%%
Now we minimize $U$ with quadrupoles $\bf{Q}$ and displacement field $\bm{d}$ which are the fundamental fields in our problem.
%%%%%%%%%%%%
\begin{align}\label{LQ6}
	\delta_{Q}U &= \delta_{Q}\int \mathcal{L} ~d^2x = \int ~d^2x \left( \Lambda_{\alpha\beta\gamma\delta}Q^{\alpha\beta} + \Gamma_{\gamma\delta}^{\alpha\beta}u_{\alpha\beta} \right) \delta Q^{\gamma\delta}, 
\end{align}
%%%%%%%%%%%
and,
%%%%%%%%%%
\begin{align}\label{LQ7}
	\delta_{d}U &= \delta_{d}\int \mathcal{L} ~d^2x = \int d^2x \left(\Sigma^{\alpha\beta}\delta u_{\alpha\beta}/2 + \sigma^{\alpha\beta}  \delta u_{\alpha\beta} + \Gamma_{\gamma\delta}^{\alpha\beta} Q^{\gamma\delta} \delta u_{\alpha\beta} \right) \nonumber \\
	&= \delta_{d}\int \mathcal{L} ~d^2x = \int d^2x \left(\Sigma^{\alpha\beta}/2 + \sigma^{\alpha\beta}  + \Gamma_{\gamma\delta}^{\alpha\beta} Q^{\gamma\delta}  \right)\delta u_{\alpha\beta}.       
\end{align}
%%%%%%%%%%
The variation with respect to $\bf Q$ gives
%%%%%%%
\begin{align}\label{LQ8}
	Q^{\alpha\beta} = -\tilde{\Lambda}^{\alpha\beta\gamma\delta}~u_{\gamma\delta}, 
\end{align}
%%%%%%%%%%%%%%%%%
where $\tilde{\Lambda}^{\alpha\beta\gamma\delta} = \Lambda^{\alpha\beta\mu\nu}\Gamma_{\mu\nu}^{\alpha\beta}$, and $\Lambda^{\alpha\beta\gamma\delta}$ is the inverse of $\Lambda_{\alpha\beta\gamma\delta}$. 
%%%%%%%%%%%%%%%%
Now we focus on Eqn.~(\ref{LQ7})  and rewrite it as, 
%see the variation with $\bm d$
%%%%%%%%%%
\begin{align}\label{LQ9}
	\delta_{d}U &=  \int d^2x \left(\Sigma^{\alpha\beta}/2 + \sigma^{\alpha\beta}  + \Gamma_{\gamma\delta}^{\alpha\beta} Q^{\gamma\delta}  \right)\delta u_{\alpha\beta} \nonumber \\
	&=  \int d^2x ~ (\Sigma^{\alpha\beta}/2 + \tilde{\sigma}^{\alpha\beta})~\delta u_{\alpha\beta},\\& \text{where}~~~~~ \tilde{\sigma}^{\alpha\beta} = \sigma^{\alpha\beta}  + \Gamma_{\gamma\delta}^{\alpha\beta} Q^{\gamma\delta}. 
\end{align}
After considering 
\begin{align}
	\label{LQ6.5}
	\delta u_{\alpha\beta} = \frac{1}{2}(\delta d_{\alpha,\beta} + \delta d_{\beta,\alpha}  )  ,
\end{align}
integrating by parts, and using $\nabla \cdot\B \Sigma$=0, we find again the equation 
\begin{align}\label{LQ10}
	\frac{\partial \tilde\sigma^{\alpha\beta}}{\partial x_{\alpha}} = 0.
\end{align}

All that happened because the quadrupole alone is being a renormalization of the elastic tensor and the background stress does not change the form of stress equilibrium equation in terms of displacement field. The effective Lagrangian and the renormalized moduli can be calculated by expressing $U$ with $Q^{\alpha\beta} = -\tilde{\Lambda}^{\alpha\beta\gamma\delta}~u_{\gamma\delta}$. Therefore, we have
%%%%%
%%%%%
\begin{align}\label{LQ11}
	\mathcal{L} &= \frac{1}{2}\Sigma^{\alpha\beta}u_{\alpha\beta}+ \frac{1}{2}~A^{\alpha\beta\gamma\delta}u_{\alpha\beta}u_{\gamma\delta} + \frac{1}{2}~\Lambda_{\alpha\beta\gamma\delta}Q^{\alpha\beta}Q^{\gamma\delta} + ~\Gamma_{\gamma\delta}^{\alpha\beta}u_{\alpha\beta}~Q^{\gamma\delta} \nonumber \\
	&= \frac{1}{2}\Sigma^{\alpha\beta}u_{\alpha\beta}+ \frac{1}{2}~A^{\alpha\beta\gamma\delta}u_{\alpha\beta}u_{\gamma\delta} + \frac{1}{2}~\Lambda_{\alpha\beta\gamma\delta} (-\tilde{\Lambda}^{\alpha\beta\mu\nu}~u_{\mu\nu}) (-\tilde{\Lambda}^{\gamma\delta\rho\sigma}~u_{\rho\sigma}) + ~\Gamma_{\gamma\delta}^{\alpha\beta}u_{\alpha\beta}~(-\tilde{\Lambda}^{\gamma\delta\rho\sigma}~u_{\rho\sigma}) \nonumber \\
	&\equiv \frac{1}{2}\Sigma^{\alpha\beta}u_{\alpha\beta}+ \frac{1}{2}\tilde{A}^{\mu\nu\rho\sigma}u_{\mu\nu}u_{\rho\sigma}, 
\end{align} 
%%%%%
where the renormalized moduli is, 
%%%%
\begin{align}\label{LQ15}
	\tilde{A}^{\mu\nu\rho\sigma} = A^{\mu\nu\rho\sigma}  + \Lambda_{\alpha\beta\gamma\delta} \tilde{\Lambda}^{\alpha\beta\mu\nu}\tilde{\Lambda}^{\gamma\delta\rho\sigma} -2\Gamma_{\gamma\delta}^{\mu\nu}\tilde{\Lambda}^{\gamma\delta\rho\sigma}. 
\end{align}

\subsection{Effect of Dipoles}
\label{Effect of Dipoles}
Now we write the energy functional with effective dipoles which are due to the gradient of high densities of quadrupoles ~\cite{21LMMPRS}. After realizing that quadrupoles are simply renormalizing the elastic tensor, we only consider the dipole interaction in brief. The Lagrangian of the system can be written as
\begin{align}
	\label{C1}
	\mathcal{L} = \frac{1}{2}\Sigma^{\alpha\beta}u_{\alpha\beta}+\frac{1}{2} A^{\mu\nu\rho\sigma}u_{\mu\nu}u_{\rho\sigma} ~+~ \frac{1}{2}\Lambda_{\alpha\beta}P^{\alpha}P^{\beta} ~+~ \Gamma_{\alpha}^{\beta}d_{\alpha}P^{\beta}.
\end{align}
%%%%%%%%
%%%%%%%
\begin{align}
	\label{C2}
	\mathcal{L} = \frac{1}{2}\Sigma^{\alpha\beta}u_{\alpha\beta}+\frac{1}{2} A^{\mu\nu\rho\sigma}u_{\mu\nu}u_{\rho\sigma} ~+~ \frac{1}{2}\Lambda_{\alpha\beta}  \partial_{\mu}Q^{\mu\alpha}  \partial_{\nu}Q^{\nu\beta}  ~+~ \Gamma_{\alpha}^{\beta}  \partial_{\mu}Q^{\mu\beta}  d_{\alpha}
\end{align}
%%%%%%%
upon minimizing with respect to the fundamental fields $\bm d$ and $\bf Q$, we find
%%%%%%%
\begin{align}
	\label{C3}
	\delta_{Q}U &= \delta_{Q}\int \mathcal{L} d^2x = \int d^2x\left( \Lambda_{\alpha\beta}P^{\alpha} + \Gamma_{\beta}^{\alpha}  d_{\alpha}  \right) \delta P^{\beta} \nonumber \\ \nonumber \\
	%%%%
	\delta_{d}U &= \delta_{d} \int \mathcal{L} d^2x = \int d^2x \left( \frac{1}{2}\Sigma^{\alpha\beta} \delta u_{\alpha\beta}+ \sigma^{\alpha\beta}\delta u_{\alpha\beta}  + \Gamma_{\beta}^{\alpha}P^{\beta}\delta d_{\alpha} \right) .
\end{align}
%%%%%%%

The first equation gives
%%%%%
\begin{align}
	\label{C4}
	\boxed{ P^{\alpha} = -\Lambda^{\alpha\beta}\Gamma_{\beta}^{\gamma} d_{\gamma} }.
\end{align}
%%%%
Now to solve the second, we use Eqn.(\ref{LQ6.5}) and perform the integration by parts to get the following equation
%%%%%%%%%%%%%
%%%%%%%
\begin{align}
	\label{C7}
	\delta_d U  &= \int d^2x ~ (\Sigma^{\alpha\beta}/2+\sigma^{\alpha\beta})\delta d_{\alpha,\beta}  + \int d^2x ~\Gamma_{\beta}^{\alpha}P^{\beta}\delta d_{\alpha} 
	\nonumber\\
	&= \oint_L dx~ (\Sigma^{\alpha\beta}/2+\sigma^{\alpha\beta})\hat{n}_\beta \delta d_\alpha + \int d^2 x ~ \left[ \Gamma_{\beta}^{\alpha}P^{\beta}  -(\sigma^{\alpha\beta})_{,\beta} \right] \delta  d_\alpha , 
\end{align}
%%%%%%%%%%
where we have used $\nabla \cdot \bf{\Sigma}=\bf{0}$ and finally get
%%%%%%%%%%
\begin{align}\label{C8}
	\boxed{(\sigma^{\alpha\beta})_{,\beta} = \Gamma_{\beta}^{\alpha} P^{\beta}}.
\end{align}
\end{widetext}
%%%%%%%%%%

After substituting the value of $P^{\alpha}$, we have
%%%%%%%%%%
\begin{align}\label{C9}
	\boxed{ (\sigma^{\alpha \beta})_{, \alpha}  = - \Gamma_{\alpha}^{\beta} \Lambda^{\alpha\nu}\Gamma_{\nu}^{\gamma}d_{\gamma} }
\end{align}
%%%%%%%%

Once we have reached the conclusion that even in the presence of a large background stress $\B \Sigma$ the quadrupoles only renormalize the elastic tensor, the addition of dipoles due to the gradients of the quadrupolar field $P^\alpha \equiv \partial_\beta Q^{\alpha\beta}$ follows verbatim the derivation provided in, say, Ref.~\cite{23CMP}. The final equation that the displacement field should satisfy 
%using Eqn.(\ref{LE5}) 
reads \cite{24FHKKP}
\begin{equation}\label{C10}
	\boxed{\Delta \B{d} + (1+ \tilde\lambda)\B \nabla (\B \nabla \cdot \B{d}) +\B \Gamma \B{d} =\B{0}},
\end{equation}
where
\begin{equation}
	\B \Gamma=
	\begin{bmatrix}
		\kappa_{e}^{2} & \mp\kappa_o^{2}  \\
		\pm\kappa_o^{2} & \kappa_{e}^{2} 
	\end{bmatrix} \ ,
	\label{odd}
\end{equation}
and $\tilde{\lambda}=\lambda/\mu$.
The $\mp$ sign of $\kappa_0^2$ means that it can be negative or positive, with an anti-symmetric counterpart $\pm \kappa_0^2$.
This equation is analytically soluble.

To solve Eqn.~(\ref{C10}) with $\B \Gamma$ as defined in Eqn.~(\ref{odd}), the displacement field $\B d$ can be separated into radial and transverse components
%%%%%%%%%%%
\begin{equation}\label{L3}
	\B {d} = d_{r}(r,\theta)\hat{r} + d_{\theta}(r,\theta)\hat{\theta}.
\end{equation}
%%%%%%%%%%%%%%%%
For concreteness we will present the solution for the case of $\kappa_o^2$ positive, and a parallel analysis can be easily done for a negative $\kappa_o^2$. We then evaluate the screening term as
\begin{equation}
	\B \Gamma \B {d}=
	\begin{bmatrix}
		(\kappa_{e}^{2}d_{r} -\kappa_o^{2}d_{\theta})~\hat{r}  \\
		(\kappa_o^{2}d_{r} + \kappa_{e}^{2}d_{\theta})~\hat{\theta} .
	\end{bmatrix} \ .
	\label{product}
\end{equation}
Now Eqn.~(\ref{C10}) can be decomposed into following coupled differential equations in $r,$ and $\theta$ (with prime and double prime standing for first or second space derivatives with respect to $r$) :
\begin{widetext}
	\begin{equation}\label{L4}
		\frac{(\tilde\lambda +2)}{r^2}\left[ r^2d_{r}^{\prime\prime} + rd_{r}^{\prime}-d_{r}   \right] -\frac{(\tilde\lambda +3)}{r^2}\frac{\partial d_{\theta}}{\partial \theta} + \frac{1}{r^2}\frac{\partial^2 d_{r}}{\partial \theta^2} + \frac{(\tilde\lambda +1)}{r} \frac{\partial^2 d_{\theta}}{\partial r\partial \theta} +\kappa_{e}^{2}d_{r} -\kappa_o^{2}d_{\theta} =0 ,
	\end{equation}
	\begin{equation}\label{L5}
		\frac{1}{r^2}\left[ r^2d_{\theta}^{\prime\prime} + rd_{\theta}^{\prime} + \frac{\partial^2 d_{\theta}}{\partial\theta^2}-d_{\theta}   \right] + \frac{(1 +\tilde\lambda)}{r^2}\left[ \frac{\partial^2 d_{\theta}}{\partial\theta^2} + r\frac{\partial^2 d_{r}}{\partial r\partial \theta} \right] + \frac{(\tilde\lambda +3)}{r^2}\frac{\partial d_{r}}{\partial \theta} + \kappa_o^{2}d_{r} + \kappa_{e}^{2}d_{\theta} =0 .
	\end{equation}
\end{widetext}

Taking into account the periodic boundary conditions on the angle $\theta$ we can now seek a solution of these equations
in the form of Fourier series for $d_{r}(r,\theta)$ and $d_{\theta}(r,\theta)$ respectively:
\begin{align}\label{L6}
	&d_{r}(r,\theta) = d_{r}(r) + \sum_{n=1}^{\infty}\left[  a_{n}(r)\cos(n\theta) + c_{n}(r)\sin(n\theta)\right] , \nonumber \\
	&d_{\theta}(r,\theta)  = d_{\theta}(r) + \sum_{n=1}^{\infty} \left[ g_{n}(r)\cos(n\theta) +e_{n}(r)\sin(n\theta)\right] .
\end{align}
%%%%%%%%%%%%
Of course, the more coefficients we keep, the more equations we need to solve. However, due to the orthogonality of the Fourier coefficients and the linearity of the equations,  different order coefficients do not mix. Thus, for simplicity, to demonstrate the decoupling,  we consider only the first-order Fourier terms, $n=1$. We have
%%%%%%%%%%%
\begin{align}\label{L7}
	&d_{r}(r,\theta) = d_{r}(r) + a_{1}(r)\cos(\theta) + c_{1}(r)\sin(\theta), \nonumber \\
	&d_{\theta}(r,\theta)  = d_{\theta}(r) + g_{1}(r)\cos(\theta) + e_{1}(r)\sin(\theta).
\end{align}

After substitution of this ansatz, and after some simplifications, Eqns.~(\ref{L4}) and (\ref{L5}) respectively take the following forms:
\begin{widetext}
	
	%%%%%%%%%%%%%%%%
	\begin{align}\label{L10}
		&\left[ r^2d_{r}^{\prime\prime}(r) + rd_{r}^{\prime}(r)-d_{r}(r)   \right] +  \frac{\kappa_{e}^{2}r^2}{(\tilde\lambda +2)}  d_{r}(r) -\frac{\kappa_{0}^{2}r^2}{(\tilde\lambda +2)}  d_{\theta}(r) \nonumber \\
		& + \left[  \left[ r^2a_{1}^{\prime\prime}(r) + ra_{1}^{\prime}(r)   \right] +  \frac{(\kappa_{e}^{2}r^2 -\tilde\lambda -3)}{(\tilde\lambda +2)} a_{1}(r) - \frac{\kappa_o^{2}r^2}{(\tilde\lambda +2)} g_{1}(r) + \frac{(\tilde\lambda+1)r}{(\tilde\lambda +2)} e_{1}^{\prime}(r) -\frac{(\tilde\lambda +3)}{(\tilde\lambda +2)}e_{1}(r)   \right] \cos(\theta) \nonumber \\
		& + \left[ \left[ r^2c_{1}^{\prime\prime}(r) + rc_{1}^{\prime}(r)   \right] + \frac{(\kappa_{e}^{2} r^2 -\tilde\lambda-3)}{(\tilde\lambda +2)} c_{1}(r) - \frac{\kappa_o^{2} r^2}{(\tilde\lambda +2)} e_{1}(r)- \frac{(\tilde\lambda+1)r}{(\tilde\lambda +2)} g_{1}^{\prime}(r) + \frac{(\tilde\lambda +3)}{(\tilde\lambda +2)}g_{1}(r)   \right] \sin(\theta) 
		=0,  
	\end{align}
	\begin{align}\label{L11}
		&\left[ r^2d_{\theta}^{\prime\prime}(r) + rd_{\theta}^{\prime}(r)-d_{\theta}(r)   \right] + \kappa_{e}^{2} r^2 d_{\theta}(r) +   \kappa_{0}^{2} r^{2}  d_{r}(r) \nonumber \\
		& + \left[  \left[ r^2g_{1}^{\prime\prime}(r) + rg_{1}^{\prime}(r)  \right]  + (\kappa_{e}^{2} r^2 -\tilde\lambda-3) g_{1}(r) +  \kappa_o^{2}r^2 a_{1}(r) + (\tilde\lambda +1)r c_{1}^{\prime}(r) +(\tilde\lambda+3) c_{1}(r)  \right] \cos(\theta) \nonumber \\
		& + \left[  \left[ r^2e_{1}^{\prime\prime}(r) + re_{1}^{\prime}(r)  \right]   + (\kappa_{e}^{2}r^2-\tilde\lambda-3) e_{1}(r) +  \kappa_o^{2}r^2 c_{1}(r) -(\tilde\lambda+1)r a_{1}^{\prime}(r) -(\tilde\lambda +3) a_{1}(r)  \right] \sin(\theta)  =0. 
	\end{align}
\end{widetext}
Since each line in the above equations has to vanish separately, these two equations produce a system of six coupled differential equations for the coefficients $d_{r}(r), d_{\theta}(r), a_1(r), b_1(r), c_1(r), e_1(r)$ and $g_1(r)$.
It is important to realize that keeping higher order terms in Eqn.~(\ref{L7}) would not change these equations, but will only add more independent equations for higher order coefficients. 

For the case of tangential rotation (radial inflation), these equations simplify further. Below we will be interested in the $r$-dependent angle average of the two components of the displacement field (cf. Eqns. (\ref{L14}) below). Angle averaging Eqns.~(\ref{L10}) and (\ref{L11}) leaves us with two coupled equations for $d_{r}(r)$ and $d_{\theta}(r)$.
The equations for $d_r(r)$ and $d_{\theta}(r)$ can be solved analytically. 
The final solutions are shown in Eqns.~(\ref{good1}) and (\ref{good2}).
%Here $\tilde \lambda=6.1$. 
%We note that in the limit $\tilde \lambda\to \infty$ the analytic solutions converge to the purely elastic solution. 
%%%%%%%%%%%%%%%%%%%%%%%%%%%%%%%%%%%%%%%%

To compare with numerical simulations, we need to extract data for the coefficients $d_r(r)$ and $d_\theta(r)$ from the measured displacement field. To this aim we compute the angle averages
\begin{eqnarray}\label{L14}
	&& d_{r}(r) = \frac{1}{2\pi}\oint_{0}^{2\pi} d_{r}(r,\theta) d\theta, \\
	&& d_{\theta}(r) = \frac{1}{2\pi}\oint_{0}^{2\pi} d_{\theta}(r,\theta) d\theta. \nonumber 
\end{eqnarray}
%%%%%%%%%%%

\section{Analytic Solutions of the Equations}
\label{anal}

In this appendix we solve the Eqns.~(\ref{L4}-\ref{L5}) for the analytical form of the radial and tangential displacements functions $d_{r}$ and $d_{\theta}$ respectively.  
%%%%%%%%%%%
We consider the following forms of the $d_{r}$ and $d_{\theta}$,
%%%%%%%%%%%
\begin{align}\label{A0}
	&d_{r}(r,\theta) = d_{r}(r)  \nonumber \\
	&d_{\theta}(r,\theta)  = d_{\theta}(r)  .
\end{align}
%%%%%%%%%%%%%%%%%
Thus Eqns.~(\ref{L4}-\ref{L5}) reduce to following coupled equations in $d_{r}(r)$ and $d_{\theta}(r)$,
%%%%%%%%%%%%%%%%%
\begin{align}\label{A00}
	[r^2d_{r}^{\prime\prime} + rd_{r}^{\prime}-d_{r} ]  +  \frac{\kappa_{e}^{2}r^2}{(\tilde\lambda +2)}  d_{r} -\frac{\kappa_o^{2}r^2}{(\tilde\lambda +2)}  d_{\theta}   & =0 ,  \nonumber \\ 
	%%%%%%%%
	[r^2d_{\theta}^{\prime\prime}  + rd_{\theta}^{\prime} -d_{\theta} ]  + \kappa_{e}^{2} r^2 d_{\theta}  +   \kappa_o^{2} r^{2} d_{r}    &=0.
\end{align}
%%%%%%%%%%%%%%
%%%%%%%%%%%%%%
We can combine the above two coupled equations in $d_{r}(r)$ and $d_{\theta}(r)$ by using two Lagrange multipliers $C_1$ and $C_2$ as follows
%%%%%%%%%%%%%%
\begin{widetext}
	%%%%%%%%%%%%%%%%%
	\begin{align}\label{A1}
		C_1 \left\lbrace [r^2d_{r}^{\prime\prime} + rd_{r}^{\prime}-d_{r} ]  +  \frac{\kappa_{e}^{2}r^2}{(\tilde\lambda +2)}  d_{r} -\frac{\kappa_o^{2}r^2}{(\tilde\lambda +2)}  d_{\theta}  \right\rbrace   
		+ C_2 \left\lbrace [r^2d_{\theta}^{\prime\prime}  + rd_{\theta}^{\prime} -d_{\theta} ]  + \kappa_{e}^{2} r^2 d_{\theta}  +   \kappa_o^{2} r^{2} d_{r} \right\rbrace  =0, 
	\end{align}
	%%%%%%%%%%%%%%
	which after some simplification can be written as,
	%%%%%%%%%%%%%%%%%%
	\begin{align}\label{A2}
		r^2\left[ C_1 d_{r}^{\prime\prime} + C_2 d_{\theta}^{\prime\prime}\right]  +   r\left[ C_1 d_{r}^{\prime} + C_2 d_{\theta}^{\prime}\right]  -  \left[ C_1 d_{r} + C_2 d_{\theta}\right]  + \left[ C_1 \frac{\kappa_{e}^{2}r^2}{(\tilde\lambda +2)} + C_2 \kappa_o^{2} r^{2} \right] d_{r} + \left[  C_2 \kappa_{e}^{2} r^2  -  C_1 \frac{\kappa_o^{2}r^2}{(\tilde\lambda +2)}   \right] d_{\theta} =0 .
	\end{align}
	%%%%%%%%%%%%%%%%
\end{widetext}
%%%%%%%%%%%%%%%
The above equation can be written in the form of Bessel differential equation. To show that this is true, we write the above equation in the following form
%%%%%%%%%%%%%%%%%%
\begin{widetext}
\begin{align}\label{A3}
	r^2\left[ C_1 d_{r}^{\prime\prime} + C_2 d_{\theta}^{\prime\prime}\right]  +   r\left[ C_1 d_{r}^{\prime} + C_2 d_{\theta}^{\prime}\right]  -  \left[ C_1 d_{r} + C_2 d_{\theta}\right] 
	+ r^2[\tilde{C_1} d_{r} + \tilde{C_2} d_{\theta}] =0 ,
\end{align}
\end{widetext}
%%%%%%%%%%%%%%%%%%%%%
where,
%%%%%%%%%%
\begin{align}\label{A4}
	& \tilde{C_1} = \left[ C_1 \frac{\kappa_{e}^{2}}{(\tilde\lambda +2)} + C_2\kappa_o^{2} \right], \nonumber \\
	& \tilde{C_2} = \left[  C_2 \kappa_{e}^{2}  -  C_1 \frac{\kappa_o^{2}}{(\tilde\lambda +2)}   \right].
\end{align}
%%%%%%%%%%

Now let us substitute 
%%%%%%%%%%%%%%%%%%
\begin{align}\label{A5}
	r^2[\tilde{C_1} d_{r} + \tilde{C_2} d_{\theta}] =  \kappa^2 r^2[ C_1  d_{r} +  C_2  d_{\theta}],
\end{align}
%%%%%%%%%%%%%%%%%%%%%
where $\kappa$ is a parameter to be obtained such that it defines the above transformation. Thus
%%%%%%%%%%%%%%%%%%
\begin{align}\label{A6}
	[\tilde{C_1} d_{r} + \tilde{C_2} d_{\theta}] =   [  \kappa^2 C_1  d_{r} +  \kappa^2 C_2  d_{\theta}],
\end{align}
%%%%%%%%%%%%%%%%%%
or,
%%%%%%%%%%%%%%%%%%
\begin{align}\label{A7}
	[\tilde{C_1} -  \kappa^2 C_1 ]d_{r} + [\tilde{C_2} -\kappa^2 C_2 ]d_{\theta} =0 .
\end{align}
%%%%%%%%%%%%%%%%%%%%%
Since, $d_{r}(r)$ and $d_{\theta}(r)$ are arbitrary, therefore to hold the above equation true, we must have 
%%%%%%%%%%%%%%%%%%
\begin{align}\label{A8}
	& \tilde{C_1} =  \kappa^2 C_1 , \nonumber \\
	& \tilde{C_2} =   \kappa^2 C_2.
\end{align}
%%%%%%%%%%%%%%%%%%%%%
Now we can solve these two equations together to determine the value of $\kappa$. After substituting the values of $ \tilde{C_1}$  and $ \tilde{C_2}$, above equations reduce to
%%%%%%%%%%%%
\begin{align}\label{A9}
	&  C_1 \frac{\kappa_{e}^{2}}{(\tilde\lambda +2)} + C_2 \kappa_o^{2}  =  \kappa^2 C_1, \nonumber \\
	&    C_2 \kappa_{e}^{2}  -  C_1 \frac{\kappa_o^{2}}{(\tilde\lambda +2)}  =   \kappa^2 C_2.
\end{align}
%%%%%%%%%%%%
With further simplifications we have
%%%%%%%%%%%%
\begin{align}\label{A10}
	&  \frac{\kappa_{e}^{2}}{(\tilde\lambda +2)} + \frac{C_2}{C_1} \kappa_o^{2}   =  \kappa^2 , \nonumber \\ \nonumber \\
	&   \kappa_{e}^{2}  - \frac{C_1}{C_2} \frac{\kappa_o^{2}}{(\tilde\lambda +2)}  =   \kappa^2 .
\end{align}
%%%%%%%%%%%%%
Let us take $\frac{C_1}{C_2} = Z$, then the above equations simplify to
%%%%%%%%%%
%%%%%%%%%%%%
\begin{align}\label{A11}
	&  \frac{\kappa_{e}^{2}}{(\tilde\lambda +2)} + \frac{1}{Z} \kappa_o^{2}   =  \kappa^2 , \nonumber \\ \nonumber \\
	&   \kappa_{e}^{2}  - Z \frac{\kappa_o^{2}}{(\tilde\lambda +2)}  =   \kappa^2 .
\end{align}
%%%%%%%%%%%%%
Now we have two equations giving the same values of $\kappa$. We equate these two equation to give a quadratic equation in $Z$,
%%%%%%%%%%%%
\begin{align}\label{A12}
	\frac{\kappa_{e}^{2}}{(\tilde\lambda +2)} + \frac{1}{Z} \kappa_o^{2}   =    \kappa_{e}^{2}  - Z \frac{\kappa_o^{2}}{(\tilde\lambda +2)},  
\end{align}
%%%%%%%%%%%%%
which simplifies to
%%%%%%%%%%%%%%%%%%%
\begin{align}\label{A13}
	Z^2 \frac{\kappa_o^{2}}{ \tilde\lambda +2 } + \left(  \frac{\kappa_{e}^{2}}{ \tilde\lambda +2 }  -  \kappa_{e}^{2}\right)  Z +  \kappa_o^{2}  =0,
\end{align}
%%%%%%%%%%%%%
or 
%%%%%%%%%%%%%%%%%%%
\begin{align}\label{A14}
	\kappa_o^{2} Z^2  -\kappa_{e}^2(\tilde\lambda + 1)  Z +   \kappa_o^{2} (\tilde\lambda + 2)  =0.
\end{align}
%%%%%%%%%%%%%
Now if we substitute $Z=\frac{C_1}{C_2}$, then the above equation can be written in terms of $C_1$ and $C_2$ as follows
%%%%%%%%%%%%%%%%%%%
\begin{align}\label{A15}
	\kappa_o^{2} C_{1}^2  -\kappa_{e}^2(\tilde\lambda + 1)  C_{1}C_{2} +   \kappa_o^{2} (\tilde\lambda + 2)C_{2}^2  =0.
\end{align}
%%%%%%%%%%%%%

The solutions of $Z$ is
%%%%%%%%%%%%%%%%%%%%%%%%%%
\begin{equation}\label{A16}
	Z = \frac{\kappa_{e}^2 (\tilde\lambda + 1) \pm \sqrt{\kappa_{e}^4 (\tilde\lambda + 1)^2 - 4(\tilde\lambda + 2)\kappa_o^4 } }{2\kappa_o^2},
\end{equation}
%%%%%%%%%%%%%%%%%%%%%%%%%%
or 
%%%%%%%%%%%%%%%%%%%%%%%%%%
\begin{equation}\label{A17}
	Z_1  = \frac{\kappa_{e}^2 (\tilde\lambda + 1) + \sqrt{\kappa_{e}^4 (\tilde\lambda + 1)^2 - 4(\tilde\lambda + 2)\kappa_o^4 } }{2\kappa_o^2}, 
\end{equation}
%%%%%%%%%%%%%%%%%%%%%%%%%%
and 
%%%%%%%%%%%%%%%%%%%%%%%%%%
\begin{equation}\label{A18}
	Z_2  = \frac{\kappa_{e}^2 (\tilde\lambda + 1) - \sqrt{\kappa_{e}^4 (\tilde\lambda + 1)^2 - 4(\tilde\lambda + 2)\kappa_o^4 } }{2\kappa_o^2}. 
\end{equation}
%%%%%%%%%%%%%%%%%%
Substituting the values of $Z$ from above equations into Eqn.(\ref{A11}), we obtain the values of $\kappa$ defining the transformation in Eqn.(\ref{A5}). Therefore, from Eqns. (\ref{A3} and \ref{A5}) we have
%%%%%%%%%%%%%%%%
\begin{widetext}
\begin{align}\label{A19}
	r^2\left[ C_1 d_{r}^{\prime\prime} + C_2 d_{\theta}^{\prime\prime}\right]  +   r\left[ C_1 d_{r}^{\prime} + C_2 d_{\theta}^{\prime}\right]  -  \left[ C_1 d_{r} + C_2 d_{\theta}\right]  + \kappa^2 r^2[C_1 d_{r} + C_2 d_{\theta}] =0 ,
\end{align}
%%%%%%%%%%%%%%%%
or 
%%%%%%%%%%%%%%%%
\begin{align}\label{A20}
	r^2\left[ C_1 d_{r}^{\prime\prime} + C_2 d_{\theta}^{\prime\prime}\right]  +   r\left[ C_1 d_{r}^{\prime} + C_2 d_{\theta}^{\prime}\right]  + (\kappa^2 r^2 -1)[C_1 d_{r} + C_2 d_{\theta}] =0.
\end{align}
%%%%%%%%%%%%%%
%%%%%%%%%%%%%%%
We can simplify it little bit more to give 
%%%%%%%%%%%%%%%%%%
\begin{align}\label{A21}
	r^2\left[ \frac{C_1}{C_2} d_{r}^{\prime\prime} +  d_{\theta}^{\prime\prime}\right]  +   r\left[ \frac{C_1}{C_2} d_{r}^{\prime} +  d_{\theta}^{\prime}\right] 
	+ (\kappa^2 r^2 -1)\left[ \frac{C_1}{C_2} d_{r} +  d_{\theta}\right]  =0 .
\end{align}
\end{widetext}
%%%%%%%%%%%%%%%%%%%%%

Now, if we substitute
%%%%%%%%%%
\begin{align}\label{A22}
	X(r) &= \frac{C_1}{C_2}d_{r}(r) + d_{\theta}(r) \nonumber \\
	&=  ~~Z d_{r}(r) + d_{\theta}(r),
\end{align}
%%%%%%%%%%%%
in the Eqn.(\ref{A21}), we obtain the following differential equation
%%%%%%%%%%%%%%%%%%
\begin{align}\label{A23}
	r^2X^{\prime\prime} + rX^{\prime}  + (\kappa^2 r^2 -1)X =0.
\end{align}
%%%%%%%%%%%%%%%%%%%%%
This is a bessel differential equation, where $\kappa$ and $Z$ are already defined above. A general solution of this equation is
%%%%%%%%%%%%%%%%%%
\begin{align}\label{A24}
	X(r) = m J_{1}(\kappa r) + n Y_{1}(\kappa r),
\end{align}
%%%%%%%%%%%%%%%%%%%%%
where $J_{1}$ and $Y_{1}$ are the Bessel functions of first kind, and the coefficients $m$ and $n$ are the constant parameters to be obtained using the boundary conditions.
Note that, we will have two solutions corresponding to the two values of $\kappa$ or $Z$.
The two values of $Z$ are $Z_1$ and $Z_2$, and the corresponding values of $\kappa$ are $\eta$ and $\zeta$ respectively.
%%%%%%%%%%%
Then we obtain following two coupled equations in $d_{r}(r)$ and $d_{\theta}(r)$ from the equations (\ref{A22}), and (\ref{A24}),
%%%%%%%%%%%%%%%%%%
\begin{align}\label{AA25}
	m_1 J_{1}(\eta r) + n_1 Y_{1}(\eta r) = Z_1 d_{r}(r) + d_{\theta}(r),
\end{align}
%%%%%%%%%%%%%%%%%%%%%
and
%%%%%%%%%%%%%%%%%%
\begin{align}\label{AA26}
	m_2 J_{1}(\zeta r) + n_2 Y_{1}(\zeta r) = Z_2 d_{r}(r) + d_{\theta}(r).
\end{align}
%%%%%%%%%%%%%%%%%%
From the above two equations we immediately obtain the analytical forms for $d_{r}(r)$, and $d_{\theta}(r)$
%%%%%%%%%%%%%%%%%%%%
\begin{widetext}
	%%%%%%%%%%%%%%%%%%
	\begin{align}\label{good1}
		d_{r}(r) = \frac{ \left[ [ m_1 J_{1}(\eta r) + n_1 Y_{1}(\eta r)] - [m_2 J_{1}(\zeta r) + n_2 Y_{1}(\zeta r)] \right] }{ Z_1 -Z_2  },
	\end{align}
	%%%%%%%%%%%%%%%%
	\begin{align}\label{good2}
		d_{\theta}(r) = \frac{ Z_2 [ m_1 J_{1}(\eta r) +  n_1 Y_{1}(\eta r)] - Z_1 [ m_2 J_{1}(\zeta r) + n_2 Y_{1}(\zeta r) ] } { \left(  Z_2 - Z_1 \right)  }.
	\end{align}
	%%%%%%%%%%%%%%%%%%%%%
\end{widetext}
%%%%%%%%%%%%%%%%%%%%%
Now we use boundary conditions to determine the coefficients $m_1, n_1$ and $m_2, n_2$. 
The boundary conditions on $d_{r}(r)$ and $d_{\theta}(r)$ are 
%%%%%%%%%%%%%%
\begin{align}\label{A29}
	d_{r}(r)|_{r=R_{\rm in}} = 0,  ~~~~~~~~~~~~ d_{r}(r)|_{r=R_{\rm out}} = 0, \nonumber \\
	d_{\theta}(r)|_{r=R_{\rm in}} = \Omega_{0},  ~~~~~~~~~~~~ d_{\theta}(r)|_{r=R_{\rm out}} = 0.
\end{align}
%%%%%%%%%%%%%
With the above boundary conditions, we have following four coupled equations in $m_1, n_1$ and $m_2, n_2$, 
%%%%%%%%%%%%%
\begin{align}
	\label{A25}
	m_1 J_{1}(\eta R_{\rm in}) + n_1 Y_{1}(\eta R_{\rm in}) &=  \Omega_{0}, ~~(i)\nonumber \\ \nonumber \\
	m_2 J_{1}(\zeta R_{\rm in}) + n_2 Y_{1}(\zeta R_{\rm in}) &=  \Omega_{0}.~~(ii)\nonumber \\  \\
	m_1 J_{1}(\eta R_{\rm out}) + n_1 Y_{1}(\eta R_{\rm out}) &= 0, ~~(iii)\nonumber \\ \nonumber \\
	m_2 J_{1}(\zeta R_{\rm out}) + n_2 Y_{1}(\zeta R_{\rm out}) &= 0.~~(iv)\nonumber \\ \nonumber  
\end{align}
%%%%%%%%%%%%
From Eqn. \ref{A25}$(i)$ and Eqn. \ref{A25}$(iii)$, we have 
%%%%%%%%%
\begin{align}\label{A31}
	\begin{bmatrix} 
		J_1(\eta R_{\rm in})& 
		Y_1(\eta R_{\rm in})\\ 
		J_1(\eta R_{\rm out})&
		Y_1(\eta R_{\rm out})
	\end{bmatrix}
	\begin{bmatrix} 
		m_1 \\
		n_1
	\end{bmatrix}
	=
	\begin{bmatrix} 
		\Omega_{0}\\ 
		0
	\end{bmatrix}.
\end{align}
%as a form of $J_{2\times 2}M_{2\times 1}=D_{2\times 1}$, where we obtain the expression of $M_{2\times 1}=J_{2\times 2} ^{-1} D_{2\times 1}$.
%%%%%%%
Thus, we get 
%%%%%%%
\begin{align}\label{A32}
	\begin{bmatrix} 
		m_1 \\
		n_1
	\end{bmatrix}
	=
	\frac{1}{\Delta_\eta}
	\begin{bmatrix} 
		Y_1(\eta R_{\rm out})& 
		-Y_1(\eta R_{\rm in})\\ 
		-J_1(\eta R_{\rm out})&
		J_1(\eta R_{\rm in})
	\end{bmatrix}
	\begin{bmatrix} 
		\Omega_{0}\\ 
		0
	\end{bmatrix},
\end{align}
where $\Delta_\eta=J_{1}(\eta R_{\rm in})  Y_{1}(\eta R_{\rm out}) -  Y_{1}(\eta R_{\rm in})J_{1}(\eta R_{\rm out}) \neq 0$.
%%%%%%%
Furthermore, we have 
%%%%%%%%%%%
\begin{align}\label{A34}
	m_1 &= \frac{Y_{1}(\eta R_{\rm out}) \Omega_{0} }{\Delta_{\eta}},\nonumber \\ \\
	n_1 &= \frac{-J_{1}(\eta R_{\rm out})  \Omega_{0}}{\Delta_{\eta}}.\nonumber 
\end{align}
%%%%%%%%%
Similarly from Eqns. \ref{A25}$(ii)$, and \ref{A25}$(iv)$, we have 
%%%%%%%
%%%%%%%
\begin{align}\label{A35}
	\begin{bmatrix} 
		J_1(\zeta R_{\rm in})& 
		Y_1(\zeta R_{\rm in})\\ 
		J_1(\zeta R_{\rm out})&
		Y_1(\zeta R_{\rm out})
	\end{bmatrix}
	\begin{bmatrix} 
		m_2 \\
		n_2
	\end{bmatrix}
	=
	\begin{bmatrix} 
		\Omega_{0}\\ 
		0
	\end{bmatrix},
\end{align}
%%%%%%%%%%%
which gives
%%%%%%%%%%%
\begin{align}\label{A36}
	\begin{bmatrix} 
		m_2 \\
		n_2
	\end{bmatrix}
	=
	\frac{1}{\Delta_\zeta}
	\begin{bmatrix} 
		Y_1(\zeta R_{\rm out})& 
		-Y_1(\zeta R_{\rm in})\\ 
		-J_1(\zeta R_{\rm out})&
		J_1(\zeta R_{\rm in})
	\end{bmatrix}
	\begin{bmatrix} 
		\Omega_{0}\\ 
		0
	\end{bmatrix}.
\end{align}
%%%%%%%%%
Furthermore, we have 
%%%%%%%%%
\begin{align}\label{A37}
	m_2 &= \frac{Y_{1}(\zeta R_{\rm out}) \Omega_{0} }{\Delta_{\zeta}},\nonumber \\ \\
	n_2 &= \frac{-J_{1}(\zeta R_{\rm out}) \Omega_{0}}{\Delta_{\zeta}}.\nonumber 
\end{align}
where $\Delta_{\zeta}=J_1(\zeta R_{\rm in}) Y_1(\zeta R_{\rm out}) - Y_1(\zeta R_{\rm in}) J_1(\zeta R_{\rm out})$.

%%%%%%%
%%%%%%%
%%%%%%
We list below all the parameters that are needed to determine the functional forms of $d_{r}(r)$ and $d_{\theta}(r)$. The values of $Z$ are
%%%%%%%%%%%%%%%%%%%%%%%%%%
\begin{align}\label{A46}
	& Z_1  =  \frac{\kappa_{e}^2 (\tilde\lambda + 1) + \sqrt{\kappa_{e}^4 (\tilde\lambda + 1)^2 - 4(\tilde\lambda + 2)\kappa_o^4 } }{2\kappa_o^2}, \nonumber \\ \nonumber \\
	& Z_2  = \frac{\kappa_{e}^2 (\tilde\lambda + 1) - \sqrt{\kappa_{e}^4 (\tilde\lambda + 1)^2 - 4(\tilde\lambda + 2)\kappa_o^4 } }{2\kappa_o^2},
\end{align}
with the first constraint as
%%%%%%%%
\begin{equation}
	(\tilde{\lambda}+1)^2\kappa_e^4-4(\tilde{\lambda}+2) \kappa_o^4 >0.
	\label{eq:1st constraint}
\end{equation}
%%%%%%%%%%%%%%%%%%%%%%%%%%
The two values of $\kappa$, i.e., $\eta$ and $\zeta$ are 
%%%%%%%%%%%%
\begin{align}\label{A47}
	&  \eta = \sqrt{ \kappa_{e}^{2}  - Z_1 \frac{\kappa_o^{2}}{(\tilde\lambda +2)}} >0, \nonumber \\ \nonumber \\
	&  \zeta = \sqrt{ \kappa_{e}^{2}  - Z_2 \frac{\kappa_o^{2}}{(\tilde\lambda +2)}} >0.
\end{align}
%%%%%%%%%%%
and the parameters $m_1, n_1, m_2, n_2$ are
%%%%%%%
\begin{align}\label{A48}
	m_1 &= \frac{Y_{1}(\eta R_{\rm out}) \Omega_{0}}
	{J_{1}(\eta R_{\rm in})  Y_{1}(\eta R_{\rm out}) -  Y_{1}(\eta R_{\rm in})J_{1}(\eta R_{\rm out})},\nonumber \\\nonumber \\
	n_1 &= \frac{-J_{1}(\eta R_{\rm out}) \Omega_{0}}
	{J_{1}(\eta R_{\rm in})  Y_{1}(\eta R_{\rm out}) -  Y_{1}(\eta R_{\rm in})J_{1}(\eta R_{\rm out})}, \nonumber \\ \\
	m_2 &= \frac{Y_{1}(\zeta R_{\rm out}) \Omega_{0} }
	{J_1(\zeta R_{\rm in}) Y_1(\zeta R_{\rm out}) - Y_1(\zeta R_{\rm in}) J_1(\zeta R_{\rm out})},\nonumber \\\nonumber \\
	n_2 &= \frac{-J_{1}(\zeta R_{\rm out}) \Omega_{0} }{J_1(\zeta R_{\rm in}) Y_1(\zeta R_{\rm out}) - Y_1(\zeta R_{\rm in}) J_1(\zeta R_{\rm out})}.\nonumber 
\end{align}

If we consider the asymptotical results for the Poisson ratio $\nu\to 0.5$ (or saying $\tilde{\lambda} \to +\infty$), i.e.,  $\kappa_{e}/\kappa_{o}\to 0^{+}$, and $\kappa \to \kappa_{e}$, from Eqn.~(\ref{A11}).
Meanwhile, Eqns.~(\ref{A48}) reduce to $m_{1}\approx m_{2}$ and $n_{1}\approx n_{2}$, and Eqn.~(\ref{A24}) to $X_{1}\approx X_{2}$. 
Thus, Eqns.~(\ref{good1}) and ~(\ref{good2}) reduce to 
\begin{equation}
	\begin{aligned}
		d_{r} &= 0, 
		\\
		d_{\theta} &=\Omega_{0}\frac{J_{1}(\kappa_{e} R_{\rm out})  Y_{1}(\kappa_{e} r) - Y_{1}(\kappa_{e} R_{\rm out})  J_{1}(\kappa_{e} r)}{Y_{1}(\kappa_{e} R_{\rm in}) J_{1}(\kappa_{e} R_{\rm out}) - Y_{1}(\kappa_{e} R_{\rm out}) J_{1}(\kappa_{e} R_{\rm in})}
		%\equiv BX_{1}+(1-B)X_{2}
		,
		\label{eq:solutions_Bessel}
	\end{aligned}
\end{equation}
as the analytical result only considering the even dipole screening.
Furthermore, without dipole screening, the result goes back to the solution of classical elastic theory (Eqn.~(\ref{LE5})) as 
\begin{equation}
	\begin{aligned}
		d_{r} &= 0, 
		\\
		d_{\theta} &= \Omega_{0} \frac{R_{\rm in}(R_{\rm out}^2 - r^2 )}{r(R_{\rm out}^2 - R_{\rm in}^2 )}.
		\label{eq:solutions_CET}
	\end{aligned}
\end{equation}

\section{Representative Examples of Tangential Displacement Fitting}
\label{Representative}

\begin{figure}[h!]
	\includegraphics[width=0.8\linewidth]{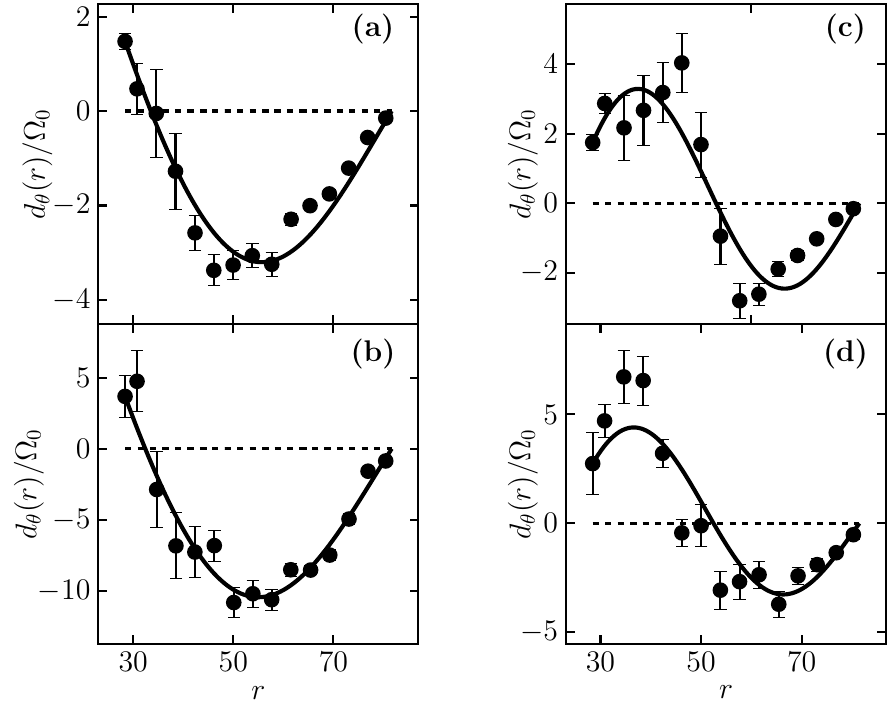}
	\caption{Other examples of the profile of the angle-averaged angular component of the displacement field $d_\theta(r)$ normalized by inner rotation $\Omega_0$ as a function of $r$, showing the reversal of particle displacement from anticlockwise to clockwise where the shear localization occurs near the boundary (Panel (a-b)) and in the bulk (Panel (c-d)). 
		The stress drops from Panel (a-d) correspond to 4.77, 19.8, 9.35, and 7.85, respectively, and fitted $\kappa_e$ corresponding to 0.067, 0.066, 0.111, and 0.109, respectively.}
	\label{loc4}
\end{figure}

In this Appendix we present additional examples of the angular-averaged angular component of the displacement field $d_\theta(r)$. In panels (a) and (b) of Fig.~\ref{loc4} the shear localization occurs near the boundary, whereas in panels (c) and (d) of Fig.~\ref{loc4} the shear localization is in the bulk. The discussion of these two possibilities, which are seen also in experiments, is provided in the main text above. 
%%%%%%%%%%%%%%%%%%%%%%%%%%%%%%%%%%%

\bibliography{ALL_citation}
%\bibliography{ALL,ALL.anomalous}
\end{document}